\documentclass[aps,prb,twocolumn,showpacs,superscriptaddress,floatfix]{revtex4-1}
\usepackage{bm}
\usepackage{amssymb}
\usepackage{graphicx}
\usepackage{amsmath}
\usepackage{dcolumn}
\usepackage[pdftex]{epsfig}
\usepackage{color}
\usepackage[german,english]{babel}
\usepackage{psfrag}
\usepackage{hyperref}
\usepackage{color}
\usepackage{cancel}

\newcommand{\ket}[1]{\ensuremath{|#1\rangle }}

\newcommand{\msu}{\uparrow}			
\newcommand{\msd}{\downarrow}		
\DeclareMathOperator{\Tr}{Tr}

\newcommand{\eV}{\ \mathrm{eV}}
\newcommand{\meV}{\ \mathrm{meV}}

\begin{document}

\title{Many-body effects on Cr(001) surfaces: An LDA+DMFT study}
\author{M. Sch\"uler}
\email{mschueler@itp.uni-bremen.de}
\author{S. Barthel}
\affiliation{Institut f{\"u}r Theoretische Physik, Universit{\"a}t Bremen, Germany}
\affiliation{Bremen Center for Computational Materials Science, Universit{\"a}t Bremen, Germany}
\author{M. Karolak}
\affiliation{Institut f{\"u}r Theoretische Physik und Astrophysik, Universit{\"a}t W{\"u}rzburg, Germany}
\author{A. I. Poteryaev}
\affiliation{M.N. Miheev Institute of Metal Physics of Ural Branch of Russian Academy of Sciences, Ekaterinburg, Russia}
\affiliation{Institute of Quantum Materials Science, Ekaterinburg, Russia}
\author{A. I. Lichtenstein}
\affiliation{Institut f{\"u}r Theoretische Physik,
Universit{\"a}t Hamburg, Germany}
\affiliation{Theoretical Physics and Applied Mathematics Department, Ural Federal University, Ekaterinburg, Russia}
\author{M. I. Katsnelson}
\affiliation{Radboud University of Nijmegen, Institute for
Molecules and Materials, The Netherlands}
\affiliation{Theoretical Physics and Applied Mathematics Department, Ural Federal University, Ekaterinburg, Russia}
\author{G. Sangiovanni}
\affiliation{Institut f{\"u}r Theoretische Physik und Astrophysik, Universit{\"a}t W{\"u}rzburg, Germany}
\author{T. O. Wehling}
\affiliation{Institut f{\"u}r Theoretische Physik, Universit{\"a}t Bremen, Germany}
\affiliation{Bremen Center for Computational Materials Science, Universit{\"a}t Bremen, Germany}

\date{\today}

\begin{abstract}
The electronic structure of the Cr(001) surface with its sharp resonance at the Fermi level is a subject of controversial debate of many experimental and theoretical works. To date, it is unclear whether the origin of this resonance is an orbital Kondo or an electron-phonon coupling effect. We have combined \textit{ab initio} density functional calculations with dynamical mean-field simulations to calculate the orbitally resolved spectral function of the Cr(001) surface. The calculated orbital character and shape of the spectrum is in agreement with data from (inverse) photoemission experiments. We find that dynamic electron correlations crucially influence the surface electronic structure and lead to a low energy resonance in the $d_{z^2}$ and $d_{xz/yz}$ orbitals. Our results help to reconvene controversial experimental results from (I)PES and STM measurements.
\end{abstract}

\maketitle
\section{Introduction}
The problem of unraveling the nature of the electronic and magnetic structure of the Cr(001) surface with its sharp resonance close to the Fermi level and strong surface magnetism has, since its first observation in angular resolved photoemission (ARPES) experiments\cite{klebanoff_observation_1984,klebanoff_investigation_1985, klebanoff_experimental_1985}, spawned many experimental and theoretical works. Inspired by scanning tunneling spectroscopy (STS) measurements\cite{stroscio_tunneling_1995} first theoretical explanations were single-particle models with the downside of having to adjust the amount of magnetic polarization of the surface to fit the experimental spectrum. New light was shed on the system by combination of STS measurements on highly clean Cr(001) surfaces and theoretical many-body techniques which suggested an orbital Kondo effect by a resonance of the degenerate $d_{xz}$ and $d_{yz}$ orbitals as the source for the resonance\cite{kolesnychenko_real-space_2002,kolesnychenko_surface_2005}. To clarify the situation, temperature dependent STS measurements have been performed but without a definite conclusion if the Kondo effect or electron-phonon effects are responsible for the electronic structure of Cr(001)\cite{hanke_temperature-dependent_2005}. Newer measurements combining STS, PES, and inverse PES (IPES)\cite{budke_surface_2008} have dealt with the symmetry properties (i.e., the orbital character) of the spectrum, revealing a $d_{z^2}$ symmetry of the supposedly Kondo peak. This speaks against an orbital Kondo scenario, as the resonance should have $d_{xz/yz}$ symmetry.  However, an alternative satisfactory explanation in terms of a single-particle theory is not at hand. The discussed electron-phonon interaction mechanisms would require coupling matrix elements for the surface of the order of 5-10 larger than the bulk value\cite{hanke_temperature-dependent_2005}. Newer PES measurements\cite{adhikary_complex_2012} show a strong temperature dependence of the resonance, namely, the emergence of a pseudogap below approximately $T=200\ \text{K}$ and the advent of a sharp resonance below $T=75\ \text{K}$. This speaks for a predominate many-body nature of the resonance and was previously not found in STS measurements\cite{hanke_temperature-dependent_2005}. We believe that both routes (single-particle theory and the simplified model for the orbital Kondo effect) have problems describing the spectrum correctly for distinctive reasons. First, the single-particle theory misses correlations effects, which cannot be negligible in a system of transition metal atoms with a considerable amount of spectral weight at the Fermi level in the paramagnetic phase\cite{allan_surface_1978}. Secondly, the model behind the orbital Kondo effect incorporates correlation effects but only deals with the $d_{xz/yz}$ orbitals and thereby misses the specific electronic structure of the system by neglecting three of the five $d$ orbitals.

Our goal is to calculate and characterize the spectrum of Cr(001) including correlation effects on the basis of a realistic description of the electronic structure. The  method of choice is the combination of density functional  theory (DFT) in the generalized gradient approximation (GGA) and dynamical mean-field theory (DMFT\cite{georges_dynamical_1996}) using the so-called LDA++ scheme\cite{anisimov_first-principles_1997,lichtenstein__1998}. The method incorporates effects stemming from strong correlations in a local approximation into the material specific framework of density functional theory. We will show that local electronic correlations play a key role for understanding the electronic structure of Cr(001). In Sec. \ref{sec:surface} we will briefly introduce the experimental spectrum which we compare our data to and review the established facts on the electronic and magnetic structure of Cr(001). In Sec. \ref{sec:ldadmft} we will give a short overview of the GGA and DMFT methods as well as the physical and numerical parameters we have used throughout this work. The results of the single-particle methods are presented in Sec. \ref{sec:res_LDA} which will be the basis for the detailed analysis of the DMFT results in Sec. \ref{sec:res_DMFT}. There, we will also prove the many-body nature of the resonance at the Fermi energy. Finally, we will give a conclusion in the last section.

\section{Models and Methods}

\subsection{Cr(001) surface}
\label{sec:surface}
Combined PES and IPES measurements make a fairly direct comparison of experimental and calculated local density of states (DOS) of the surface possible. The most up-to-date combined PES \textit{and} IPES measurements\cite{budke_surface_2008} obtained at a temperature of $T=295\ \text{K}$ ($\beta\approx39.3\eV^{-1}$) reveal three main features in the spectrum (reproduced in Fig. \ref{fig:expDOS} (b)). First, a broad peak at $-0.5\eV$ dominated by $d_{xz/yz}$ character. Secondly, a sharp peak around the Fermi level at $0.0\eV\pm 0.05\eV$ which is mainly of $d_{z^2}$ character. Thirdly, a broad peak at $1.65\eV$ which is again dominated by $d_{xz/yz}$ character.

\begin{figure}[htb]
\begin{center}
\includegraphics[width=1\columnwidth]{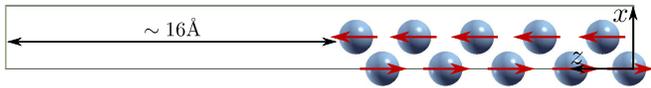}
\end{center}
\caption{(Color online) Unit cell used to model the Cr(001) surface. The $z$ coordinates of the two outermost atoms per surface were relaxed in GGA. Arrows indicate the magnetic ordering. The figure is generated with XCrySDen\cite{kokalj_computer_2003}.}
\label{fig:unitcell}
\end{figure}
In order to reproduce the structure of the spectrum realistically, we model the Cr(001) surface by a slab of ten Cr atoms and a region of vacuum of approximately $16\text{\AA}$ stacked in $z$ direction as depicted in Fig. \ref{fig:unitcell}. This slab is periodically continued in $x$ and $y$ direction. Earlier DFT calculations\cite{habibi_electronic_2013,bihlmayer_electronic_2000} have shown that a slab of ten atoms is sufficiently large so that the innermost atoms behave like bulk atoms. We calculate the spectra near the experimental temperature ($T\approx290 \ \text{K}$, $\beta=40\eV^{-1}$). We analyze the temperature dependence of the spectrum by simulating one lower ($\beta=60\eV^{-1}$) and one higher ($\beta=20\eV^{-1}$) temperature. The magnetic ground state of the Cr(001) surface can be viewed as a stack of ferromagnetic Cr layers with antiferromagnetically alternating magnetization between neighboring layers\cite{habibi_electronic_2013} (see Fig. \ref{fig:unitcell}). The cubic symmetry present in the bulk, leading to three degenerate $t_{2g}$ and two $e_{g}$ $d$ orbitals, is lifted at the surface leaving only the $d_{xz}$ and $d_{yz}$ orbitals degenerate.

\subsection{LDA+DMFT}
\label{sec:ldadmft}
For the density functional part of the calculations we use the projector augmented-wave method\cite{kresse_ultrasoft_1999} implemented in the Vienna Ab initio Simulation Package (\textsc{vasp}). We have taken account of structural changes in the vicinity of the surface by optimizing the coordinate perpendicular to the surface of the two uppermost and lowermost atoms in GGA. To perform many-body calculations on top of the paramagnetic GGA calculations, a projection of delocalized low-energy Kohn-Sham states to a localized basis is done. We choose the 70 Kohn-Sham states lowest in energy for this task which capture the spectral weight of the relevant hybridized $s$, $p$, and $d$ states. The subspace in which Coulomb interaction is treated by DMFT is defined by the five $d$ orbitals on each Cr atom. The projection on atomlike $d$-orbital wave functions are readily available from \textsc{vasp} standard output within the PAW-framework and are post processed to form an orthonormal basis\cite{amadon_plane-wave_2008}, also referred to as PLO(V)\cite{karolak_general_2011}. Finally, the 20 spare states form the uncorrelated subspace and are defined as the orthogonal complement of the correlated states which we calculate by a Gram-Schmidt process. The projection yields a single-particle Hamilton matrix for each in-plane wave number $\mathbf{k}$: $H(\mathbf{k})^{ij}_{\alpha\beta}=\langle i,\alpha | H(\mathbf{k})|j,\beta\rangle$ in the localized basis $\ket{i,\alpha}$, labeled by a layer index $i$ and a combined index $\alpha$ for $d$ orbitals and states of the orthogonal complement. The correlated states are augmented with a local Coulomb interaction defined by
\begin{align}
U_{\alpha\beta\gamma\delta}^i = \sum_{k=0}^{2l} a_k(\alpha_m\beta_m,\gamma_m\delta_m)F^k_i,
\end{align}
where $a_k(\alpha_m\beta_m,\gamma_m\delta_m)$ are proportional to products of Gaunt coefficients \cite{slater_quantum_1960,eder_robert_multiplets_2012}, Greek indices are atomic orbitals, $i$ is the layer index, $l=2$ is the angular momentum quantum number for $3d$ electrons in the case of Cr, and $F^0_i = U_i$, $F^2_i = 14/(1+0.625)J_i$ and $F^4_i=0.625F^2_i$ are Slater parameters. The average Coulomb interaction $U_i$ and Hunds exchange interaction $J_i$ are taken from \textit{ab initio} constrained random-phase approximation (cRPA\cite{aryasetiawan_frequency-dependent_2004}) calculations\cite{csacsiouglu_strength_2012}. The Coulomb interaction has a strong layer dependence\footnote{We have checked the sensitivity of our results with respect to the used Coulomb matrix elements by using constant average interaction of $U_i=4.95\eV$ (the bulk value from the reference) and $U_i=2.0\eV$ with $J_i=0.9$ from Ref. [\onlinecite{chioncel_textitab_2003}]. The prior parameters lead to basically the same spectra as the layer dependent ones. The considerably smaller parameters lead to no resonance at the Fermi level.} ($U_{1/10}=3.44\eV$, $U_{2/9}=4.64\eV$, $U_{3/8}=4.73\eV$, $U_{4/7}=4.94\eV$, and $U_{5/6}=4.95\eV$) while the Hunds exchange interaction is constant ($J_i=0.65\eV$). On first sight it seems counterintuitive that the Coulomb interaction is smaller at the surface than in the bulk, as screening should be less efficient due to a smaller screening volume. However, surface electronic structure effects such as the appearance of surface states and effective band narrowing can also increase screening at the surface. In the case of the Cr(001) surface, the electronic structure effect dominates the volume effect, thus leading to a decrease of the surface interaction matrix elements with respect to the bulk value, as discussed in detail in Ref. [\onlinecite{csacsiouglu_strength_2012}]. We neglect all Coulomb-interaction terms besides density-density terms. We solve the multi-orbital Hubbard model of the slab of ten atoms by multi-site DMFT which allows for spatially inhomogeneous Coulomb interaction and antiferromagnetic ordering\cite{georges_dynamical_1996}. Compare, e.g., Refs. [\onlinecite{lechermann_towards_2015}] and [\onlinecite{valli_dynamical_2010}] for a similar approach. In multi-site DMFT the lattice Green function reads
\begin{align}
\begin{split}
[G&(i\omega_n,\mathbf{k})^{-1}]^{ij}_{\alpha\beta} =\\
& \left[\left(i\omega_n + \mu \right)\delta_{\alpha\beta}  - \Sigma(i\omega_n)^i_{\alpha \beta} \right]\delta_{ij}- H(\mathbf{k})^{ij}_{\alpha \beta},
\end{split}
\end{align}
where the DMFT approximation of a local, i.e., $\mathbf{k}$-independent, self-energy is apparent. The double-counting term is absorbed into the self-energy.  The local lattice Green function $G(i\omega_n)$ is obtained by $\mathbf{k}$-averaging $G(i\omega_n,\mathbf{k})$. We compute the Weiss field,
\begin{align}
[\mathcal{G}^0(i\omega_n)^{-1}]^i_{\alpha\beta} =[G(i\omega_n)^{-1}]^i_{\alpha\beta} + \Sigma(i\omega_n)^i_{\alpha \beta}, \label{eq:weiss}
\end{align}
for each Cr atom and solve the resulting effective impurity problems for each Cr atom until self-consistency in the DMFT loop is reached. 
To solve the impurity problems we use the continuous-time quantum Monte Carlo (CTQMC) algorithm in the hybridization expansion\cite{werner_hybridization_2006,werner_continuous-time_2006} implemented in the w2dynamics software package\cite{parragh_conserved_2012}. We cope with the double-counting problem, which is inherent to LDA++ approaches, by the requirement that the total occupation on each impurity obtained from the DMFT Green function $G$ matches the corresponding occupation obtained from the bath Green function. This is called trace double-counting correction and reads
\begin{align}
\Tr \rho_{\alpha\beta}^\text{imp} = \Tr \rho_{\alpha\beta}^{0,\text{loc}},
\label{eq:trace}
\end{align}
where $\rho$ is the density matrix for each atom. This leads to satisfactory results in metallic systems\cite{karolak_electronic_2013}. To assess the influence of the choice of the double-counting energy we have used an alternative double-counting scheme, namely enforcing 4.5 electrons on each Cr atom\footnote{We replace the right hand side of Eq. (\ref{eq:trace}) with 4.5. In contrast, the trace double counting leads to 4.7 electrons in the case of the surface atom.}. For the analytic continuation from imaginary to real-frequency spectral functions we use the maximum entropy method\cite{bryan_maximum_1990,silver_maximum-entropy_1990} as implemented in the w2dynamics software package. The antiferromagnetic order is achieved by starting with a constant shift of the real part of the self-energy in the first iteration of the DMFT loop. We stress the importance of global spin flips of configurations in the Markov chains in the presence of magnetic polarization in order to prevent getting stuck in local minima. We calculate the magnetic moment for atom $i$ by\cite{georges_dynamical_1996}
\begin{align}
m_i = \frac{1}{\beta}  \sum_{n,\alpha}  e^{i\omega_n0^+}\left( G_\msu(i\omega_n)^i_{\alpha\alpha} - G_\msd(i\omega_n)^i_{\alpha\alpha} \right).
\end{align}
After reasonable convergence of the self-energy, some final iterations with high statistics ($10^5-10^6$ measurements depending on the inverse temperature) are done. The autocorrelation times for each inverse temperature are estimated by the ratio of the acceptance rate of inserting operators in the trace and the position of the maximum in the expansion histogram\cite{nicolaus_parragh_strongly_2013} leading to $N_\text{corr}=4500,14000,40000$ for $\beta=(20,40,60)\eV^{-1}$, respectively.
\section{Results}
\subsection{DFT spectra}
\label{sec:res_LDA}
\begin{figure}[htb]
\begin{center}
\mbox{
\includegraphics[width=1\columnwidth]{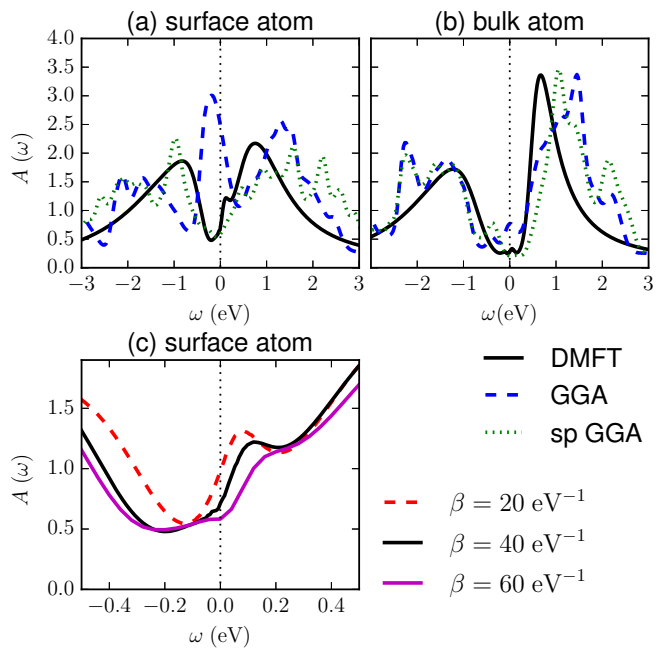}
}
\end{center}
\caption{(Color online) Orbital and spin averaged local density of states (LDOS) from LDA+DMFT simulations at the temperature $\beta=40\eV^{-1}$ (solid line), GGA (dashed lines), and spin-polarized GGA (dotted lines) calculations. (a) shows the LDOS at the surface atom. (b) shows the LDOS of the center atom. (c) shows details near the Fermi level of DMFT spectra for different temperatures, $\beta=20\eV^{-1}$ (dashed line), $\beta=40\eV^{-1}$ (solid black line), and $\beta=60\eV^{-1}$ (solid magenta line).}
\label{fig:totalDOS}
\end{figure}

\begin{figure}[htb]
\begin{center}
\mbox{
\includegraphics[width=1\columnwidth]{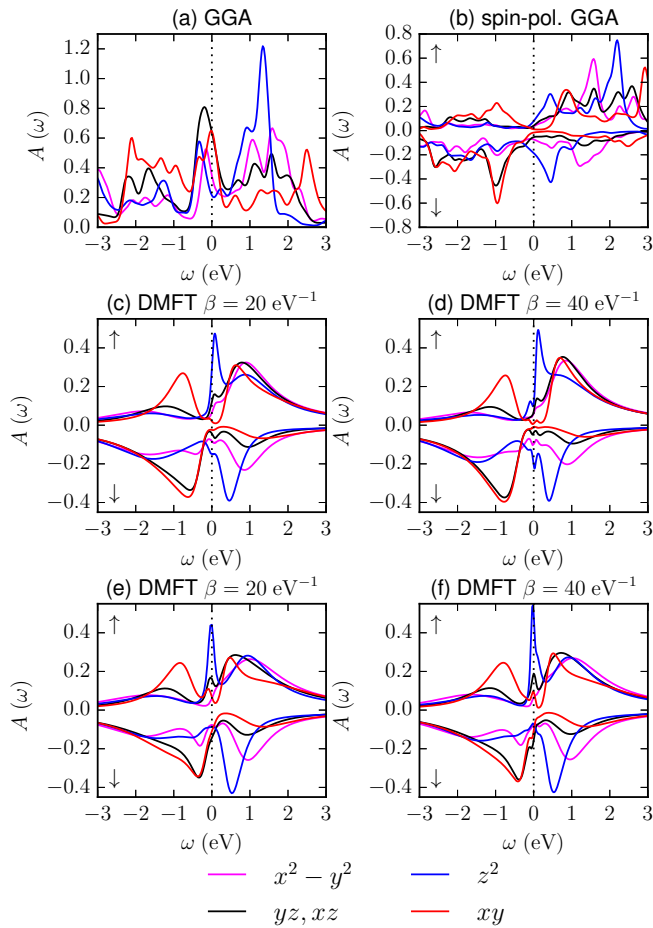}
}
\end{center}
\caption{(Color online) Orbitally resolved local density of states of the surface atom from (a) GGA, (b) spin-polarized GGA, and  LDA+DMFT simulations at different double-counting energies and temperatures (c) $\beta=20\eV^{-1}$ and (d) $\beta=40\eV^{-1}$ at $E_\text{dc}\approx13.5\eV$ (trace double counting) and (e) $\beta=20\eV^{-1}$ and (f) $\beta=40\eV^{-1}$ at $E_\text{dc}\approx12.2\eV$.}
\label{fig:orbDOS}
\end{figure}

The broadened local density of states for the paramagnetic GGA calculation are shown in Figs. \ref{fig:totalDOS} for (a) a surface atom and (b) an atom of the fifth layer, which is henceforth named bulk atom. The results are very close to reported tight-binding calculations\cite{allan_surface_1978}. At first glance, the structure of the spectrum resembles the experimental one, with three peaks at about the right positions. However, neglecting the magnetic order and correlation effects leads to many differences in detail: The peak at the Fermi energy is far too big and broad. The orbital characters shown in Fig. \ref{fig:orbDOS} (a) do not coincide with the experimental data reported in Ref. [\onlinecite{budke_surface_2008}]. The peak at the Fermi energy has nearly evenly divided orbital weight among all $d$ orbitals and the upper broad peak is mainly of $d_{z^2}$ character. The peak at the Fermi energy is absent in the bulk case. The high spectral weight at the Fermi energy together with a considerable Coulomb interaction ($U_1=3.44\eV$) is a possible basis for a large magnetic moment and strong many-body effects at the surface.

The static mean-field treatment of spin polarization effects in the case of the spin-polarized GGA calculations leads to a splitting of the states close to the Fermi energy. The position of the broad peaks is estimated to be $-1.0\eV$ and $1.6\eV$, which corresponds roughly to the experimental positions ($-0.5\eV$ and $1.65\eV$). The large initial (``paramagnetic'') peak at the Fermi energy leads to a larger magnetization at the surface ($|m|\sim2.31\mu_B$) than for the bulk atoms ($|m|\sim 1.25\mu_B$). The orbitally resolved density of states shown in Fig. \ref{fig:orbDOS} (b) reveals that the lower broad peak is of mainly $d_{xz,yz}$ and $d_{xy}$ character which is in line with the experimental findings. The upper peak has also $d_{xz,yz}$ and $d_{xy}$ character but considerable $d_{x^2-y^2}$ weight is also found. Instead of a resonance directly at the Fermi level, spin polarized GGA yields a feature of $d_{z^2}$ character $0.5\eV$ above the Fermi energy. In summary, spin-polarized GGA is able to describe the broad peaks above and below the Fermi energy reasonably well, whereas the sharp feature at the Fermi energy is absent.

\subsection{DMFT spectra}
\label{sec:res_DMFT}
For all three investigated temperatures we have found antiferromagnetic solutions in DMFT. Similar to the spin-polarized GGA calculation we find a larger magnetic moment for the surface atoms ($|m|\sim2.2\mu_B$, $2.4\mu_B$ and $3.0\mu_B$) than for the bulk atoms ($|m|\sim 1.5\mu_B$, $1.8\mu_B$ and $2.5\mu_B$) for $\beta=20\eV^{-1}$, $40\eV^{-1}$ and $60\eV^{-1}$, respectively. As in the case of spin-polarized GGA, this results from the larger paramagnetic GGA density of states at the Fermi energy (c.f. Figs. \ref{fig:totalDOS} (a) and (b)) and is therefore a consequence of the rearrangement of the electronic states at the surface. The magnetic moment shows a strong dependence on the temperature.

\begin{figure}[htb]
\begin{center}
\mbox{
\includegraphics[width=0.7\columnwidth]{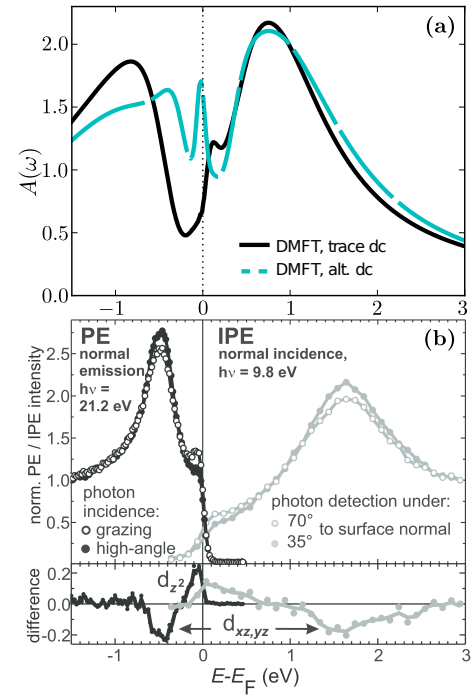}
}
\end{center}
\caption{(Color online) (a) DMFT surface spectra at $\beta=40\eV^{-1}$ with two different double-counting schemes: trace double counting (solid line) and an alternative scheme (see text, dashed line). (b) Experimental (I)PES data reproduced with authorization from M. Donath and M. Bode from Ref. [\onlinecite{budke_surface_2008}]. PE and IPE spectra obtained at $T = 295\ \text{K}$ in experimental geometries with different sensitivities to $d_{z^2}$-like (open dots) and
$d_{xz,yz}$-like (filled dots) orbital character. For details see Ref. [\onlinecite{budke_surface_2008}]. }
\label{fig:expDOS}
\end{figure}

We first analyze the density of states obtained using the trace double-counting scheme: The local density of states at $\beta=40\eV^{-1}$ summed over the $d$ orbitals and spins for the surface atom is shown in comparison to the GGA spectra in Fig. \ref{fig:totalDOS} (a). The same is shown for a bulk atom in Fig. \ref{fig:totalDOS} (b). A direct comparison of the experimental data and the DMFT spectra of the surface atom is given in Fig. \ref{fig:expDOS}. For the surface atom we can observe a three-peak structure resembling the experimental situation. In particular, the lower and upper peak are located at $-0.8\eV$ and $0.75\eV$, respectively. The peak at the Fermi level is shown in detail in Fig. \ref{fig:totalDOS} (c) for all investigated temperatures. It is located at about $0.1\eV$ for $\beta=20\eV^{-1}$ and $\beta=40\eV^{-1}$. The peak is shifted towards $150\meV$ above the Fermi level and appears as a shoulder for $\beta=60\eV^{-1}$. The resonance is slightly sharper for the simulation at $\beta=20\eV^{-1}$. While the position of the central peak fits nicely to the experiment, the position of the lower and upper peak is off by $\sim0.3\eV$ and $\sim1\eV$, respectively.

To test the influence of the choice of the double-counting energy we have calculated spectra using the alternative double-counting scheme described above. For the surface atom, the double counting we arrive at is $1.3\eV$ smaller as in the case of the trace double counting ($E_\text{dc}^\text{trace}\approx13.5\eV$ and $E_\text{dc}^\text{4.5}\approx12.2\eV$). For comparison, using a fully localized limit approach\cite{anisimov_first-principles_1997} we arrive at $E_\text{dc}^\text{FLL}\approx13.2\eV$. The resulting total density of states for the surface atom at $\beta=40\eV^{-1}$ is presented in Fig. \ref{fig:expDOS} (a). The resonance at the Fermi energy is more pronounced and shifted by approximately $100\meV$ to the Fermi energy. The lower broad peak is shifted to considerably higher energy ($-0.4\eV$) and lies nearly on top of the corresponding experimental peak ($-0.5\eV$). The position of the upper broad peak is not affected by the smaller double-counting energy. 

A possible reason for the discrepancy between experimental and calculated spectral peak positions, especially of the upper broad peak for both double-counting schemes, could be the neglect of Coulomb interaction effects (particularly exchange interaction) between the correlated and uncorrelated bands and those between the uncorrelated bands themselves, leading to a too small overall exchange splitting. Considering the lower peak and the resonance at the Fermi energy, the total spectrum from the smaller double-counting energy coincides better with the experimental data. The bulk spectra do not change dramatically with the temperature and are very close to the spin-polarized GGA treatment: dynamic correlation effects seem to be less important far from the surface\footnote{From the analytically continued spectrum only it is hard to assess whether or not differences between static- and dynamical-mean field, as the ones discussed in Ref. [\onlinecite{sangiovanni_static_2006}] are present or not.}, which is in agreement with Ref. [\onlinecite{chioncel_textitab_2003}]. Indeed, the spectrum for the second layer already shows nearly no peak at the Fermi level (not shown).

The orbital characters of the spectra obtained with the trace double counting for the two temperatures showing the resonance are presented in Fig. \ref{fig:orbDOS} (c) and (d). The broad peaks show very similar characters to the spin-polarized GGA calculations: the lower peak is mainly of $d_{xz/yz}$ and $d_{xy}$ character, the upper peak is mainly of $d_{xz/yz}$, $d_{xy}$, $d_{x^2-y^2}$, and slightly of $d_{z^2}$ character. The sharp central peak is dominated by a feature in the minority spin channel with $d_{z^2}$ character but also carries some spectral weight from the other $d$ orbitals, particularly the $d_{xz}$ and $d_{yz}$ orbitals. For $\beta=40\eV^{-1}$ the $d_{z^2}$ orbital also shows a sharp resonance in the majority spin channel.

The orbitally resolved spectral functions for the surface atom in the case of the alternative double-counting scheme are shown in Figs. \ref{fig:orbDOS} (e) and (f) for the same temperatures as before. The overall character of the peaks is unaffected by the change of the double-counting energy. The $t_{2g}$ contribution of the lower peak is sharper than in the former case and shifted towards the Fermi level. The smaller double-counting energy leads to additional minor $d_{xy}$ character of the resonance. The resonance in the $d_{z^2}$ majority spin part of the spectrum for the trace double counting present at $\beta=20\eV^{-1}$ and $\beta=40\eV^{-1}$ is absent in this case. In conclusion, for both double-counting energies the orbital character of all peaks are in line with the experimental data reported in Ref. [\onlinecite{budke_surface_2008}].

The comparison of the spectra at different temperatures shows that the resonance at the Fermi level tends to shift from $\beta=40\eV^{-1}$ on. This shift is contradictory to temperature dependent STS measurements\cite{hanke_temperature-dependent_2005} of the resonance which show a resonance for temperatures from $\beta=33.5\eV^{-1}$ to $\beta=527\eV^{-1}$, which narrows with smaller temperature without any shift. This behavior in the simulations could be an artifact of the density-density approximation of the Coulomb tensor and has to be investigated further. However, recent temperature dependent PES measurements\cite{adhikary_complex_2012} do find an emerging pseudogap above $T=200 \ \text{K}$ ($\beta=58\eV^{-1}$), which is in qualitative agreement with the results of our simulations. The experimental data also show a sharp feature reappearing in the spectrum for temperatures below $T=50\ \text{K}$; a temperature unfortunately computationally too expensive for the QMC method used in this work. In summary, the orbital character of all three peaks found in the experimental data could be confirmed by the DMFT simulations.

%

Kondo-like resonances originate from degenerate levels in contact with a bath. In the original sense of the Kondo effect, these levels are the spin levels. One characteristic property of Kondo-type resonances is their behavior under breaking of this degeneracy by an energy $\Delta$: below a critical splitting on the scale of the Kondo temperature $T_K$ the resonance is pinned to the Fermi energy. Above the critical energy the resonance is split by approximately $\Delta$ and broadened\cite{costi_kondo_2000}. In the case of Cr(001), the suggested \textit{orbital} Kondo effect in Ref. [\onlinecite{kolesnychenko_real-space_2002}] results from the degenerate $d_{xz}$ and $d_{yz}$ orbitals. 

To check if the $d_{xz/yz}$ contribution of the above found resonance is of a Kondo-like nature, we test its robustness against a small splitting of the orbital energy, i.e., an artificial crystal field. Similarly, we test the many-body nature of the $d_{z^2}$ contribution of the resonance by shifting the $d_{z^2}$ orbital energy. To this end, we investigate five-orbital Anderson impurity models with artificially broken crystal symmetries. The impurity models are defined by the hybridization function of the surface atom found in the DMFT self-consistency for the trace double counting, the appropriate chemical potential and Coulomb matrix. The corresponding impurity actions for the surface atom ($i=1$) thus read

\begin{align}
\begin{split}
S_\text{imp} = &-\sum_{\alpha\beta}\int d\tau  d\tau'  d^\dagger_\alpha(\tau)\left([\mathcal{G}_{0}^{-1}(\tau - \tau') ]_{\alpha\beta}\right.\\
&\left. + \delta_{\alpha \beta} \delta(\tau-\tau') \Delta_\alpha \right) d_\beta(\tau')\\
&+\sum_{\alpha\beta\gamma\delta}U_{\alpha\beta\gamma\delta}^1 \int d\tau  d_\alpha^\dagger(\tau) d_\beta^\dagger(\tau) d_\gamma(\tau)d_\delta(\tau),
\end{split}
\end{align}

where $\mathcal{G}_{0}$ is the Weiss field of the surface atom found in the self-consistent solution of Eq. (\ref{eq:weiss}). If not further specified, the additional orbital dependent crystal field $\Delta_\alpha$ is zero.

\begin{figure}[tb]
\begin{center}
\mbox{
\includegraphics[width=1\columnwidth]{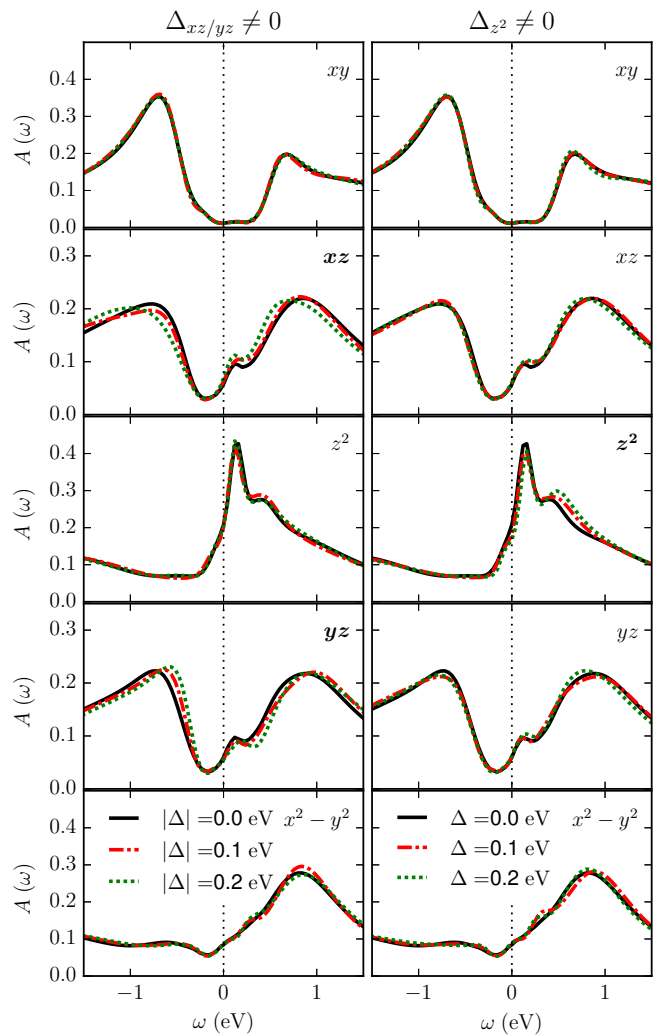}
}
\end{center}
\caption{(Color online) Spin averaged local density of states of the (from top to bottom) $d_{xy}$, $d_{xz}$, $d_{z^2}$, $d_{yz}$, and $d_{x^2-y^2}$ orbitals of the surface atom obtained from the solution of an AIM with the self-consistent hybridization function from DMFT at $\beta=40\eV^{-1}$ and additional crystal field splitting of $\Delta_{xz}=0,-0.1,-0.2\eV$ and $\Delta_{yz}=0,0.1,0.2\eV$ with full, dashed and dotted lines (left panels). The same is shown in the right panels with a crystal field of $\Delta_{z^2}=0,0.1,0.2\eV$.}
\label{fig:shift}
\end{figure}

We first break the $d_{xz/yz}$ degeneracy by simultaneously applying a negative crystal field to the $d_{xz}$ orbital and a positive to the $d_{yz}$ orbital. The strength of the crystal field is $|\Delta_{xz/yz}|=0.1\eV$ and $|\Delta_{xz/yz}|=0.2\eV$. The resulting spectra of all orbitals are depicted in the left panels of Fig. \ref{fig:shift}. First, the spectra of the orbitals with unaltered crystal field (i.e., $d_{xy,z^2,x^2-y^2}$) are basically unaffected by the additional crystal field $\Delta_{xz/yz}$.  While the broad peaks of the $d_{xz/yz}$ above and below the Fermi level experience a rigid shift in the direction of the respective crystal field, revealing them as single-particle peaks, the position of the central resonance is basically not changed. This speaks very much for a many-body nature of the resonance, as expected from the prior comparison of the GGA and DMFT spectra. We further analyze the resonance by shifting the $d_{z^2}$ orbital by $\Delta_{z^2}=0.1\eV$ and $\Delta_{z^2}=0.2\eV$. The spectral contributions of all orbitals are shown in the right panels of Fig. \ref{fig:shift}. Again, the spectra of all other orbitals (i.e., $d_{xy,xz,yz,x^2-y^2}$) are basically unaffected by applying $\Delta_{z^2}$.  The central resonance in the $d_{z^2}$ contribution does not change with the crystal field, speaking for a many-body effect as the source of the resonance. In contrast, the maximum at $\sim 0.5\eV$ is rigidly shifted, revealing it as a single-particle feature.

The analysis of the resonance's behavior subject to a shift of single-particle energies rules out a pure single particle nature of the resonance. Its robustness is an indicator for a many-body nature. However, a conventional Kondo effect does not likely explain the resonance because the temperature dependence found in this work contradicts the expectations from the Kondo effect (sharpening of the resonance with lower temperatures; splitting under symmetry breaking) and the corresponding Kondo temperature would be extremely high ($T_K\gtrsim300\ \text{K}$). The nature of the resonance seems to lay in the combination of the rearrangement of the electronic structure at the Cr(001) surface (massively more states at the Fermi energy in contrast to bulk Cr) and the multi-orbital interaction effects, especially in the $d_{z^2}$ and $d_{xz/yz}$ orbitals. The interaction introduces two major effects: first, the spin dependent splitting of mainly the $d_{xz/yz}$ orbitals and, secondly, the appearance of quasiparticle peaks in $d_{z^2}$ and $d_{xz/yz}$ orbitals. We conclude that the resonance is a complex many-body effect in the $d_{z^2}$ and $d_{xz/yz}$ orbitals due to dynamic local-correlation effects. The low-energy resonance discussed here for temperatures $T>190\ \text{K}$ compares best to the low-energy resonance observed in (I)PES in Ref. [\onlinecite{budke_surface_2008}] and displays a clear temperature dependence in our calculations. The relation of this resonance to the temperature dependent low-energy resonance reported in STM studies\cite{kolesnychenko_real-space_2002,hanke_temperature-dependent_2005} remains however unclear and would require calculations at low temperatures.

\section{Conclusion}

We have used the LDA+DMFT method to calculate the spectral function of the Cr(001) surface and have compared our results with experimental data on the spectral function at different temperatures. We have derived and chosen the parameters of the model based on \textit{ab initio} calculations. We could identify the main experimental features in our data and have found their orbital character to coincide with experimental findings. By the comparison of spin-polarized GGA and DMFT calculations and an analysis of the DMFT spectrum by an artificial crystal field we could show that the resonance at the Fermi level is a many-body feature of mainly $d_{z^2}$ and slightly $d_{xz/yz}$ character. We could thus show that dynamic local-correlation effects play a key role to the electronic nature of the Cr(001) surface. The shift of the resonance to higher energies for inverse temperatures above $\beta=40 \eV^{-1}$ is in line with PES measurements but contradictory to STS measurements and its clarification needs more investigations.

\textit{Acknowledgments.} The authors thank Matthias Bode for inspiring discussions. M.I.K. acknowledges financial support by ERC Advanced Grant No. 338957 and by NWO via Spinoza Prize. A.I.P. acknowledges the state assignment of FASO of Russia (theme “Electron” No. 01201463326). This work has been funded by the Deutsche Forschungsgemeinschaft through the research units FOR 1162 (MK) and FOR 1346 (G.S., S.B., M.S., A.I.L., and T.O.W.).  Computer time at the HLRN (Project ``hbp00030'') is acknowledged.

\bibliographystyle{apsrev4-1}
\bibliography{newbib}

\end{document}